\documentclass{an}
\usepackage{graphicx}
\usepackage{times}
\def\lya{\mbox{Ly$\alpha$}}
\def\ecs{\mbox{~erg~cm$^{-2}$~s$^{-1}$}}
\def\lessim{\mathrel{\hbox{\rlap{\hbox{\lower4pt\hbox{$\sim$}}}\hbox{$<$}}}}
\def\gtrsim{\mathrel{\hbox{\rlap{\hbox{\lower4pt\hbox{$\sim$}}}\hbox{$>$}}}}

\Volume{00}              
\Year{0000}              
\Month{00}               
\Pagespan{000}{000}      

\begin{document}
\lhead[\thepage]{L. Christensen et al.: Integral field observations of
  DLA galaxies} 
\rhead[Astron. Nachr./AN~{\bf 325} (2004) 2]{\thepage}
\headnote{Astron. Nachr./AN {\bf 325} (2004) 2, 124--127}

\title{Integral field observations of Damped Lyman-$\alpha$ Galaxies}

\author{L.~Christensen\inst{1}, S.~F.~S\'anchez\inst{1}, K.~Jahnke\inst{1},
  T.~Becker\inst{1}, A.~Kelz\inst{1}, L.~Wisotzki\inst{1,2}, and   M.~M.~Roth\inst{1}}
\institute{Astrophysikalisches Institut Potsdam, An der Sternwarte 16, 
14482 Potsdam, Germany
\and 
Universit\"at Potsdam, Am Neuen Palais 10, 14469 Potsdam, Germany}
\date{Received {date will be inserted by the editor}; 
accepted {date will be inserted by the editor}} 

\abstract{We report preliminary results from a targeted investigation on
  quasars containing damped Lyman-$\alpha$\ absorption (DLA) lines as well
  strong metal absorption lines, carried out with the Potsdam Multi Aperture
  Spectrophotometer (PMAS).  We search for line-emitting objects at the same
  redshift as the absorption lines and close to the line of sight of the QSOs.
  We have observed and detected the already confirmed absorbing galaxies in
  Q2233+131 ($z_{abs}=3.15$) and Q0151+045 ($z_{abs}$=0.160), while failing to
  find spectral signatures for the $z=0.091$ absorber in Q0738+313. From the
  Q2233+131 DLA galaxy, we have detected extended \lya\ emission from an area
  of 3\arcsec$\times$5\arcsec.  
 }

\correspondence{lchristensen@aip.de}

\maketitle

\section{Introduction}
Some quasars exhibit strong Lyman-$\alpha$ absorption lines in their spectra,
which originate in neutral gas clouds located in the line of sight towards the
QSOs. If the column density of neutral gas is larger than 2$\times10^{20}$
cm$^{-2}$ the line is termed a Damped Lyman-$\alpha$ (DLA) line. It has been
shown that these systems contain a large bulk of neutral gas compared to the
total baryonic mass in the Universe (Storrie-Lombardi \& Wolfe 2000).
Metallicities of the systems investigated through high resolution spectroscopy
of the metal absorption lines associated with DLA lines give [Fe/H] values
between 0.01 and 1 times the solar value. These results, along with the fact
that the column density is typical for giant molecular clouds, have lead to
the hypothesis that DLAs are associated with galaxies and are star forming
objects.  Many investigations have been carried out in order to find the
parent galaxies of the DLA systems, but only 6 cases at $z\gtrsim2$ have
spectroscopic confirmation, while at lower redshifts 11 systems have been
confirmed (Chen \& Lanzetta 2003). The investigations have tried to answer the
questions of whether are large rotating galaxies (Wolfe at al. 1986) or gas
rich dwarfs (Hunstead et al. 1990), which is supported by the hierarchical
merging scenario (Haehnelt, Steinmetz and Rauch 1998).

Observations of the absorbing galaxies at lower redshifts ($z\lessim1$) have
shown that DLA galaxies have luminosities in the range $0.02<L/L^*<1.2$ (Rao
et al.  2003) and have diverse morphological types in agreement with
observations in Le~Brun at al. (1997). At higher redshifts ($z>1.9$) the few
confirmed DLA galaxies are compact and fainter than an $L^*$ galaxy (M{\o}ller
et al.  2002).
 
The standard way of confirming the parent DLA galaxy is to perform deep
broad-band or narrow-band imaging of the fields of QSOs containing DLAs in
order to find candidate DLA galaxies close to the line of sight (Warren,
M{\o}ller, \& Jakobsen 2001; Le Brun et al. 1997).  These candidate galaxies
need confirmation by follow-up spectroscopy in order to determine their
redshifts.  Confirming the absorbing galaxy is therefore conventionally a time
consuming two step procedure, which can be avoided with integral field
spectroscopy (IFS).

In these proceedings we describe an on-going project to study DLAs with
integral field spectroscopy, carried out with the PMAS
spectrograph\footnote{Webpage
  http://www.aip.de/groups/opti/pmas/OptI\_pmas.html}, the preliminary data
reduction, and some early results.

\section{Project description}
The general idea behind this investigation is to analyse the advantages of
using IFS for detecting and confirming the DLA galaxies. We have divided the
investigation into two parts, one for the high redshift ($z\gtrsim1.9$), the
other for the lower redshift DLAs. This partition comes in naturally, in the
sense of which emission line we want to detect. For the high-$z$ galaxies, the
\lya\ line is redshifted into the optical region, while for the lower redshift
objects other emission lines originating in star forming regions are used,
typically [\ion{O}{II}], H$\beta$, or [\ion{O}{III}].

First of all, we observed a few already confirmed DLA galaxies and one strong
Mg II absorption system in order to investigate the capabilities of the PMAS
instrument for this task, while planning the investigations of systems for
which no DLA galaxies have been detected previously.

For simplicity the lower redshift DLA galaxies are brighter, providing ideas
of what can be expected for the high redshift DLA galaxies.  An evolution of
the systems with redshift is expected since the average metallicities of DLAs
were shown to increase with decreasing redshift (Prochaska et al. 2003). We do
not plan to investigate the metallicities of the systems, but focus on the
emission properties and velocity structure of the galaxies.

\section{Data reduction}
The observations are carried out with PMAS mounted on the 3.5m telescope at
Calar Alto. The PMAS spectrograph has 256 fibers coupled to a 16$\times$16
lens array, each lens covering 0\farcs5 on the sky on a side (Roth et al.
2000). The preliminary data reduction proceeds as follows: First the bias
level is subtracted, and the location of the 256 spectra are found using
spectra of a continuum lamp. Knowing the position of the spectra on the CCD,
the data are extracted and the wavelength calibration is done using an
exposure of HgNe line lamps.  Flat-fielding is done using a twilight sky
spectrum using an average transmission of each fiber. The flux calibration is
done the standard way by comparing the reduced spectra with that of
spectrophotometric stars (Becker 2002).

\section{Observations of DLA galaxies}
In this section we present observations of three QSOs which have strong
absorption lines in their spectra. Two of the ones presented have galaxies at
the same redshifts which have previously been confirmed spectroscopically by
other groups using long slit spectroscopy.

\subsection{Q0151+045}
\begin{figure}
\resizebox{\hsize}{!}
{\includegraphics[bb=0 0 415 415,clip]{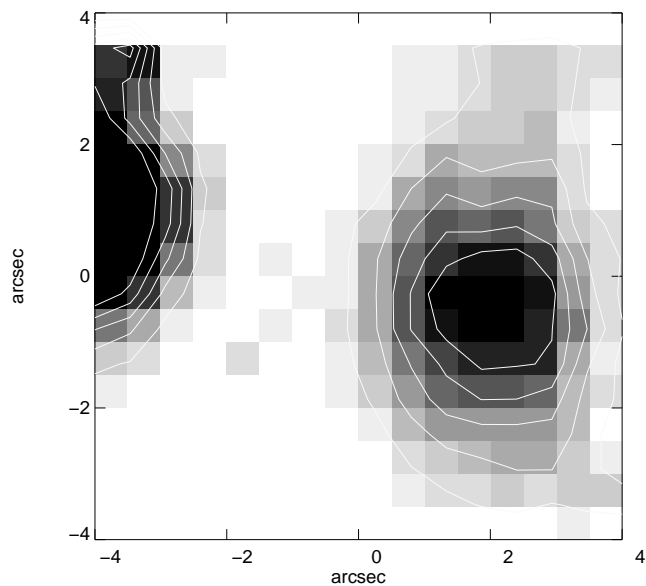}}
\caption{Negative 8\arcsec$\times$8\arcsec\ image of the region surrounding
  Q0151+045. North is up and east is left, and the QSO is located at the edge
  of the field towards the east. This QSO has a strong \ion{Mg}{II} absorber at
  $z=0.160$, but no DLA. The absorption line is caused by a galaxy
  $\sim$6\arcsec\ to the west of the QSO.}
\label{fig:q0151_im}
\end{figure}

This QSO at $z=0.404$ has a \ion{Mg}{II} absorption system at $z=0.160$ with a
very high equivalent width of 1.55{\AA}. With such a large EW it is suspected
that this system belongs to a DLA line, but a UV spectrum has shown that the
column density of the absorber is below the DLA limit (Rao \& Turnshek 2000).
The galaxy responsible for the \ion{Mg}{II} absoption is readily detected; it
is a bright $R=19$ galaxy at an impact parameter of 6\arcsec\ (Bergeron et al.
1988).  Given the brightness of the galaxy it posed as a good test case for
the analysis of the low redshift absorbers.  The observations were done with
PMAS in August 2002, for a total of 2 hours integrations, with the QSO placed
at the edge of the field of view. This was done in order to observe the
absorbing galaxy simultaneously, and secondly allowing tracing of the
differential atmospheric refraction in the 3D data-cubes, which is done more
accurately for point sources.  We corrected the data cubes for a substantial
differential atmospheric refraction as the airmass during the 4 integrations
varied between 1.3 and 1.6. An image of the galaxy is shown in
Fig.~\ref{fig:q0151_im}.

\begin{figure}
\resizebox{\hsize}{!}
{\includegraphics[bb=0 180 337 700,clip,angle=90]{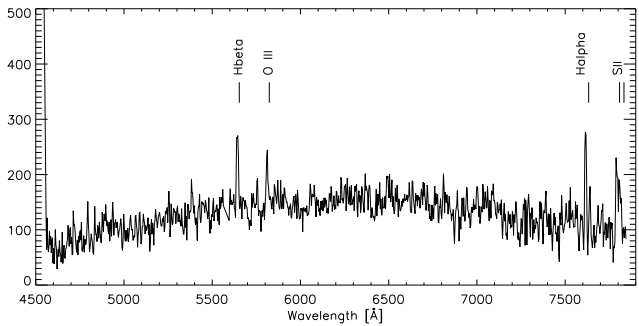}}
\caption{Spectrum of the galaxy responsible for the \ion{Mg}{II} absorption 
  lines in the spectrum of Q0151+045. Clearly visible are emission lines 
  from H$\alpha$, H$\beta$, [\ion{O}{III}], and [\ion{S}{II}].}
\label{fig:q0151_spec}
\end{figure}

\begin{figure*}
\begin{tabular}{ccc}
\hspace*{2.5cm} \large{H$\alpha$} & \hspace*{4.8cm}\large{H$\beta$} &
\hspace*{4.6cm} \large{[O} \normalsize{III}\large{]} \\
\vspace*{-1cm}
\end{tabular}\\
\begin{minipage}[c]{.33\textwidth}
\resizebox{\hsize}{!}
{\includegraphics[bb=0 0 415 415,clip]{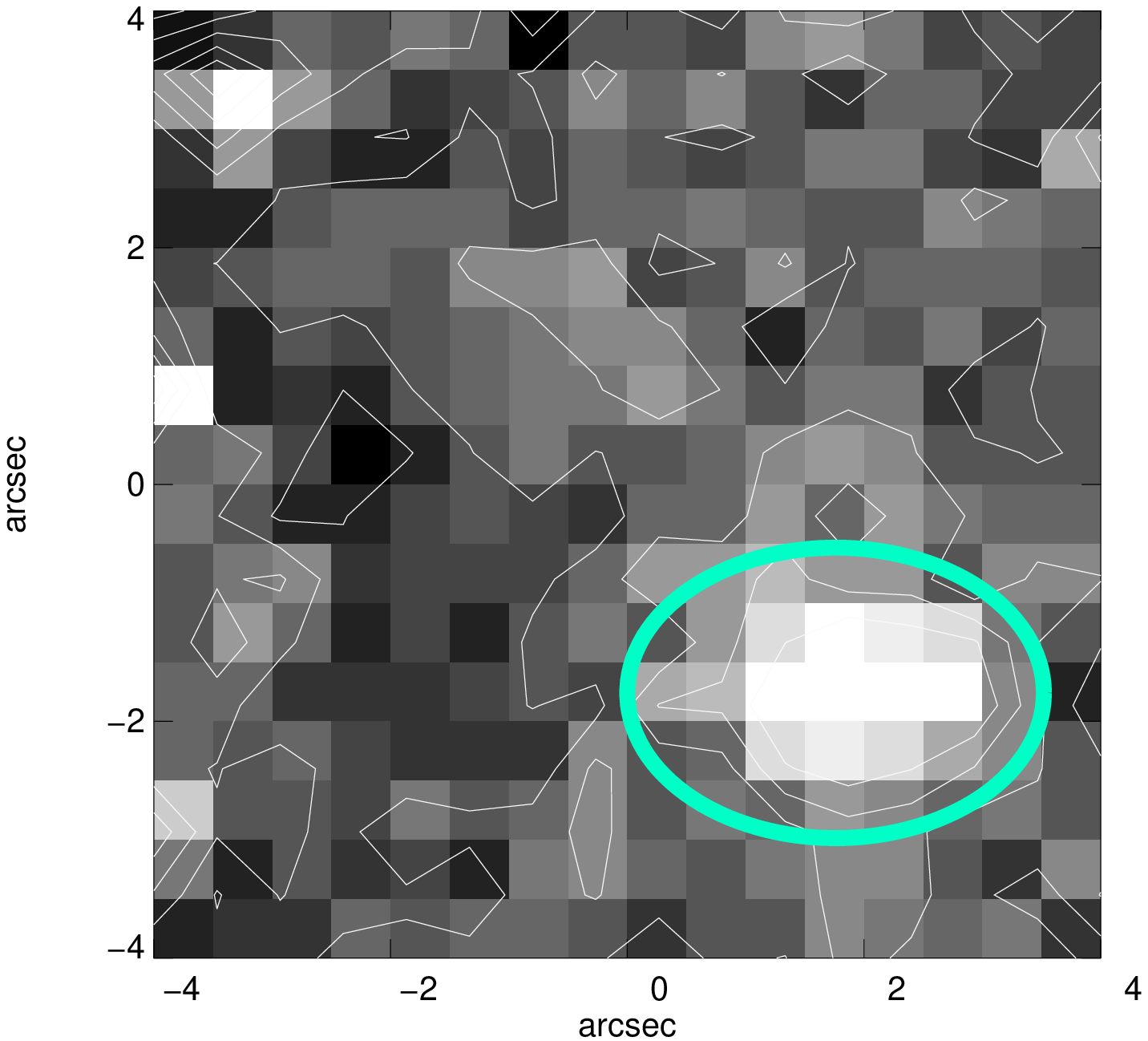}}
\end{minipage}%
\begin{minipage}[c]{.33\textwidth}
\resizebox{\hsize}{!}
{\includegraphics[bb=0 0 415 415,clip]{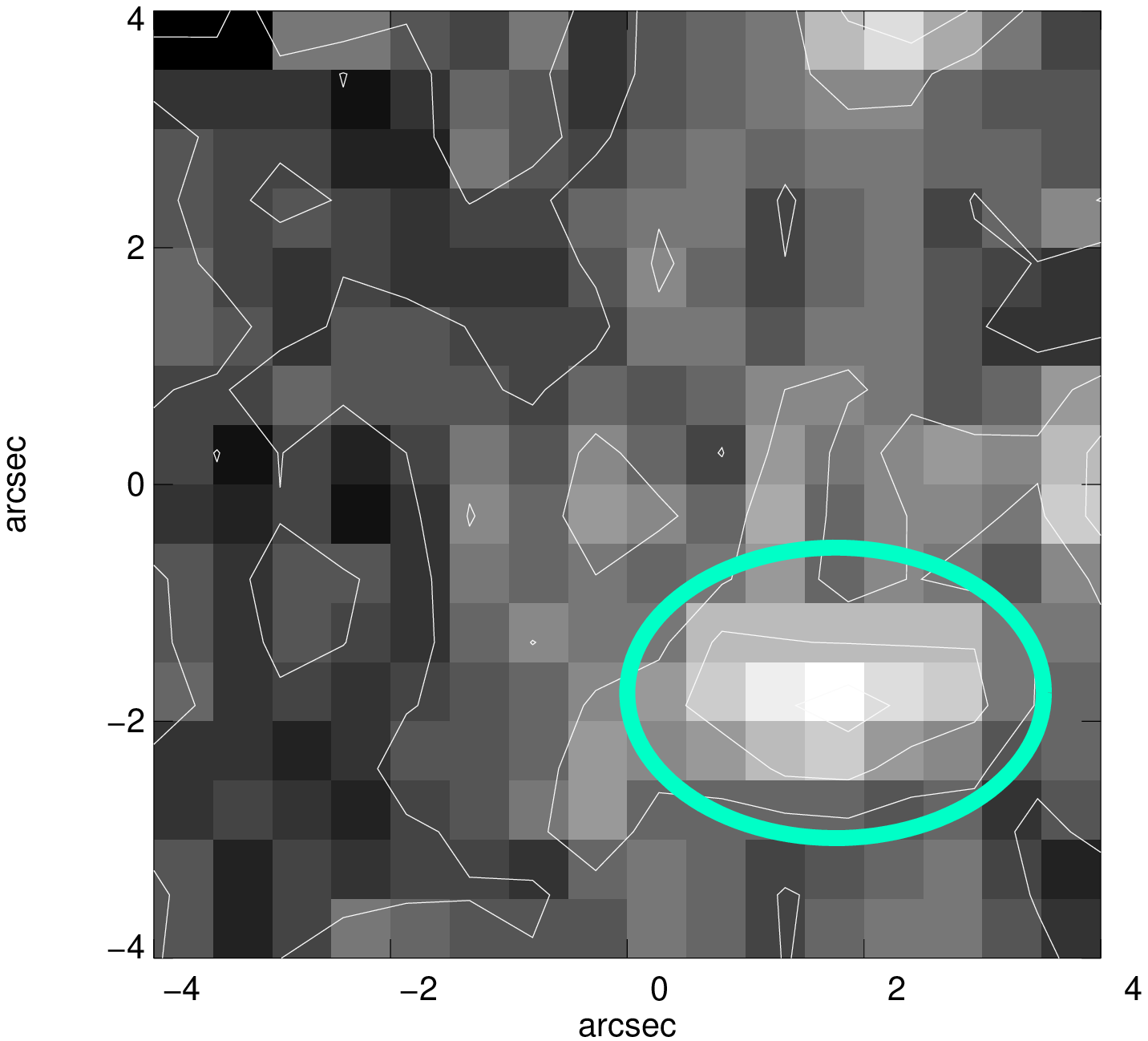}}
\end{minipage}%
\begin{minipage}[c]{.33\textwidth}
\resizebox{\hsize}{!}
{\includegraphics[bb=0 0 415 415,clip]{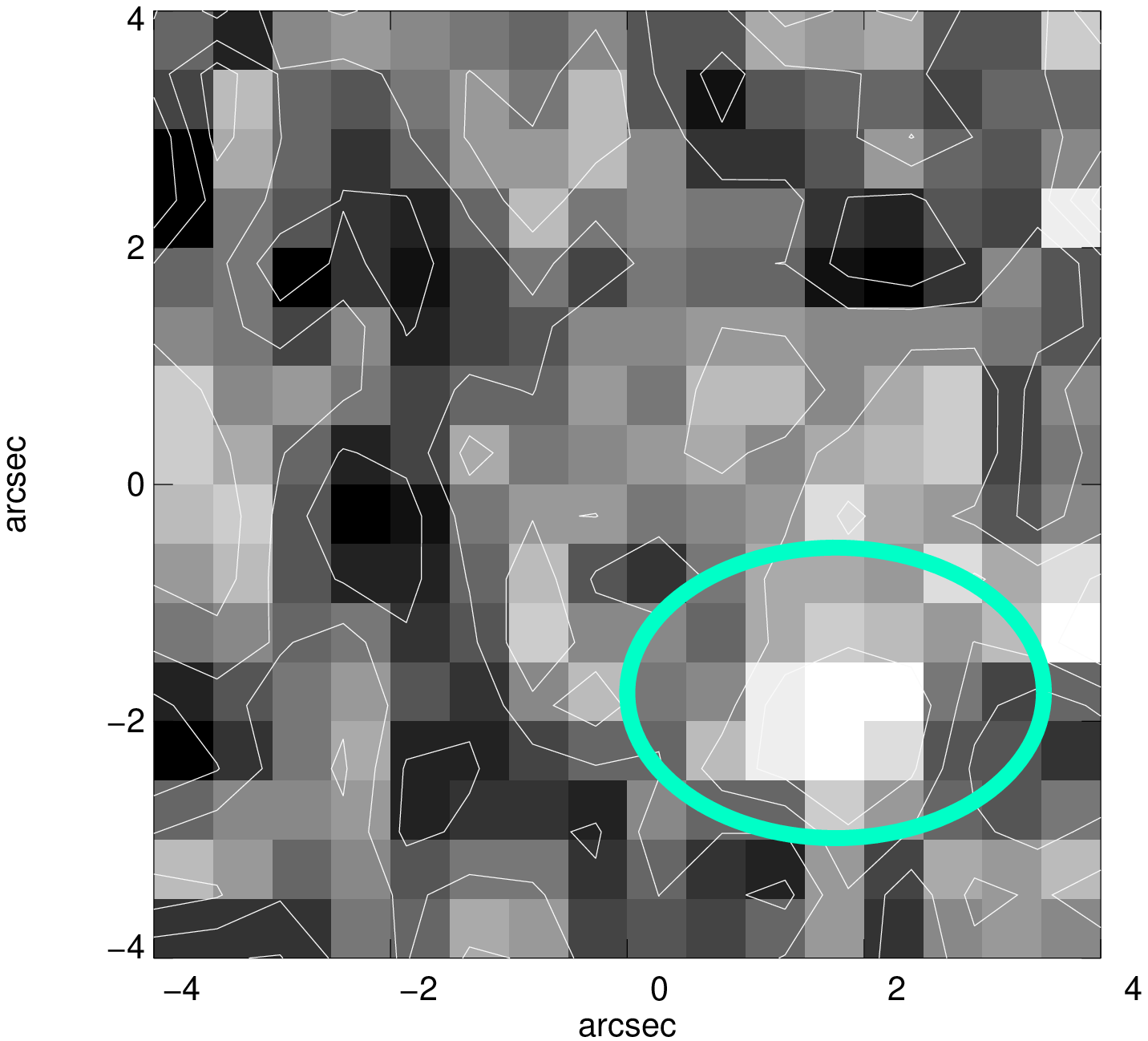}}
\end{minipage}
\caption{Images of the Q0151+045 absorber at the wavelengths containing the
  emission lines H$\alpha$, H$\beta$, and [\ion{O}{III}], respectively. North
  is up and east is left. In all cases the continuum has been subtracted. The
  three panels show that the line emission from the galaxy is located to the
  south of the center.}
\label{fig:q0151_ha}
\end{figure*}

Summing the spatial pixels (spaxels) belonging to the absorbing galaxy, after
background subtraction, yields the spectrum shown in
Fig.~\ref{fig:q0151_spec}. Emission lines from H$\beta$, [\ion{O}{III}],
H$\alpha$, and [\ion{S}{II}] are clearly seen.  By subtracting the continuum
near the emission lines, one can produce maps showing the extension of
individual emission line regions. Examples for H$\alpha$, H$\beta$, and
[\ion{O}{III}] are shown in Fig.~\ref{fig:q0151_ha}, where one sees that the
emission occurs mostly in the southern part of the galaxy indicated by the
ellipses. Without the correction for differential atmospheric refraction, the
locations of the emission will change with wavelengths, which will have great
implications for deriving the spaxel to spaxel flux ratios of e.g.
H$\alpha$/H$\beta$.

\subsection{Q0738+313}
This QSO at $z=0.630$ has two DLA lines at $z=0.0912$ and $z=0.2212$,
respectively. The higher redshift DLA galaxy has been confirmed as a dwarf
galaxy at an impact parameter of 5\farcs7 (Cohen 2001), while the lower
redshift has not been identified despite many efforts (Turnshek et al. 2001).
It is suspected that the DLA galaxy is a low surface brightness galaxy having
a very small impact parameter (essentially lying in front of the QSO). These
features make spectroscopic confirmation very complicated using traditional
slit spectroscopy, but with IFS it is possible to sum up the spectra in an
extended area on the sky increasing the signal detected for the low surface
brightness galaxy. With this in mind we have obtained data of this object with
PMAS, but have not been able to identify any spectral features in the
preliminary data reduction.

\subsection{Q2233+131}
\begin{figure}
\resizebox{\hsize}{!}{\includegraphics[angle=0]{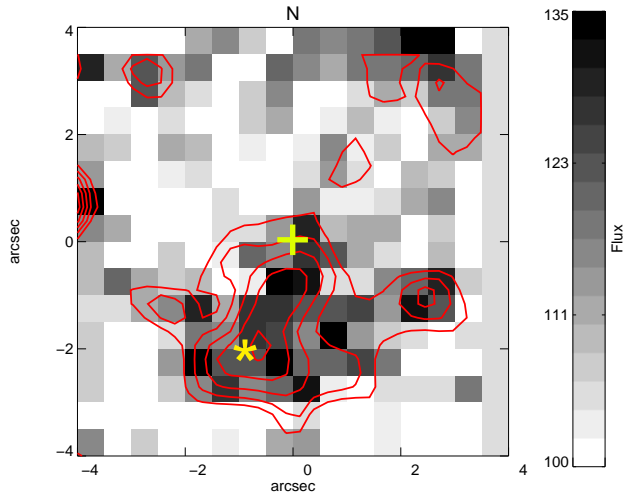}}
\caption{Narrow band image of the Q2233+131 at $\lambda$=5040--5058 {\AA}
  corresponding to the DLA line in the QSO spectrum, i.e. in the region where
  the light from the QSO has been absorbed. The location of the QSO is at the
  center of the field at (0,0) indicated by the '+'. One sees that an object
  with an extension of 3\arcsec$\times$5\arcsec is present.  Some
  contamination at of flux from the QSO at $\lambda> 5055$ {\AA} is present,
  and is caused by including emission from the damped wing of the QSO. In the
  two dimensional spectra the \lya\ emission is well separated from the QSO.
  The location of the parent galaxy from STIS observations (M{\o}ller at al.
  2002) is indicated by the '*'. } 
\label{fig:map}
\end{figure}

\begin{figure}
\resizebox{\hsize}{!}
{\includegraphics[bb=0 180 337 700,clip,angle=90]{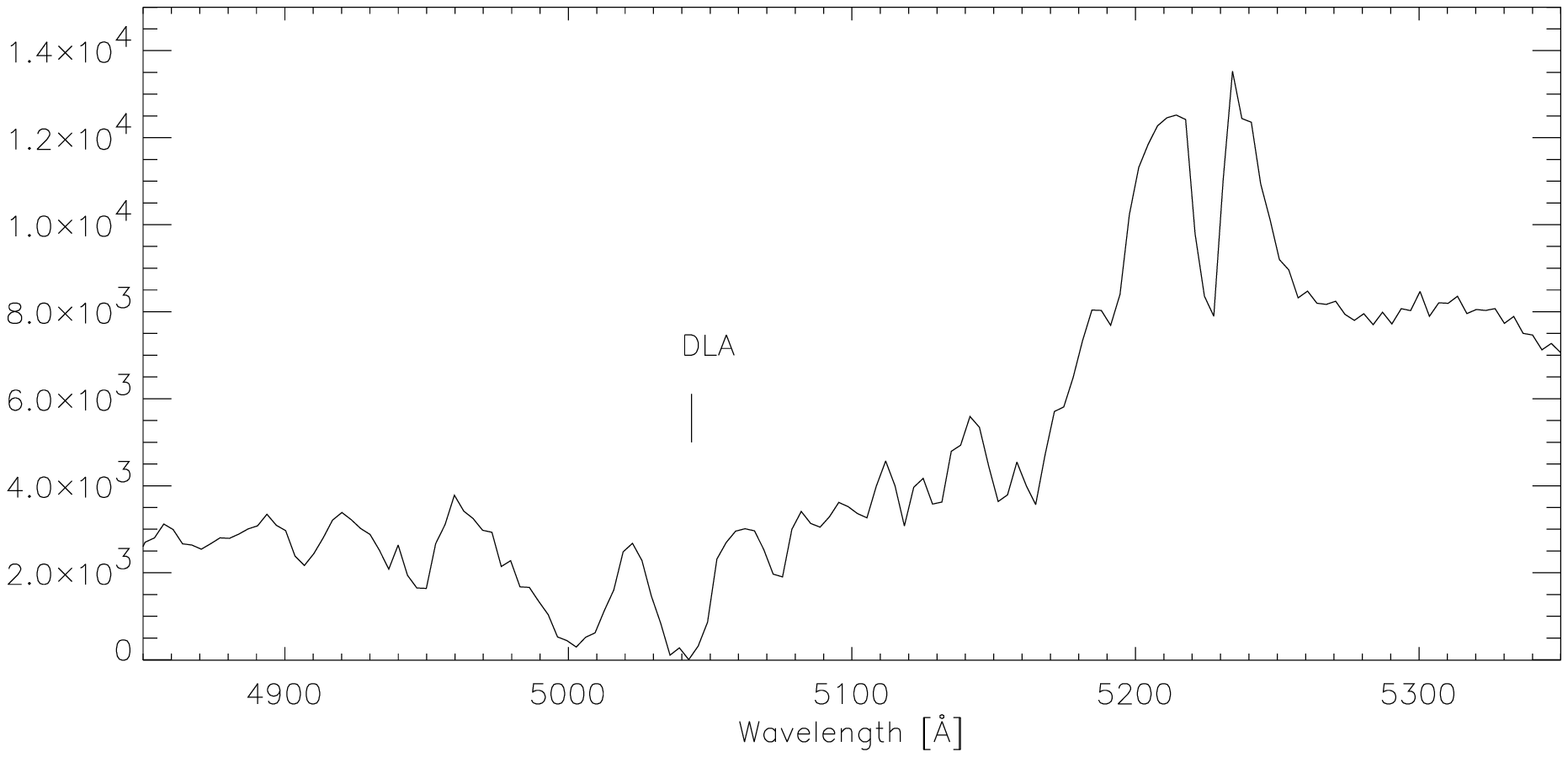}}
\resizebox{\hsize}{!}
{\includegraphics[bb=54 180 319 700,clip,angle=90]{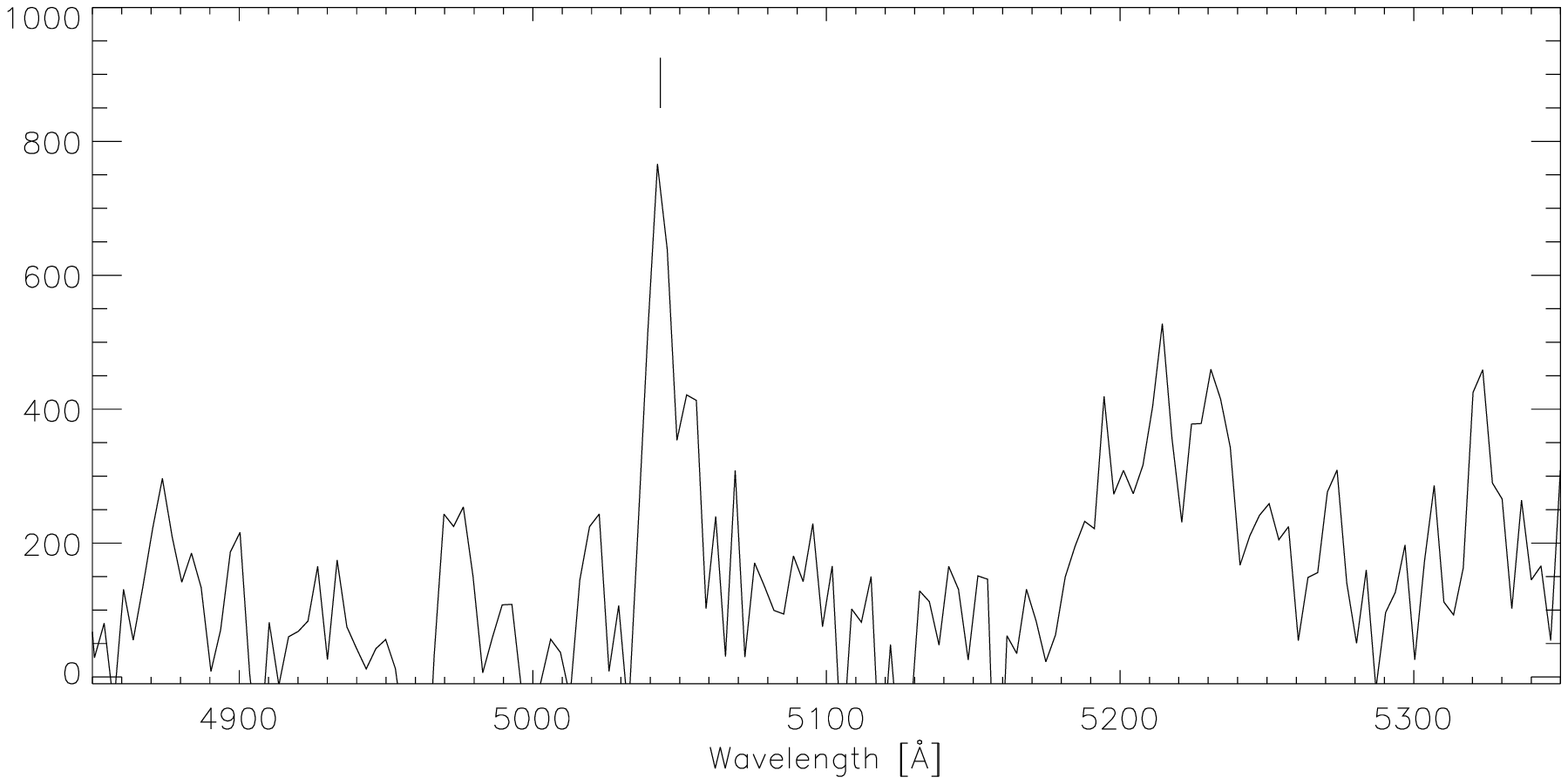}}
\caption{Spectrum of Q2233+131  around the DLA
  line. Selecting the spaxels which appear to be associated with the extended
  nebula in Fig.~\ref{fig:map}, results in the spectrum presented in the lower
  panel. The nebula has \lya\ emission at the same redshift as the DLA line in
  the QSO spectrum with a velocity difference of $\sim$300 km~s$^{-1}$.}
\label{fig:spec}
\end{figure}

This QSO at $z=3.29$ has a broad \lya-absorption line from a $z=3.15$ neutral
cloud with a column density of $N_{\mathrm{H~I}}=10^{20}$~cm$^{-2}$, i.e.
lower than the classical limit of DLAs. Djorgovski et al. (1996) identified
the galaxy responsible for the absorption line at an impact parameter of
2\farcs5 using long slit spectroscopy with the Keck telescope. This distance
corresponds to 17 kpc in a flat $\Lambda$ dominated cosmology. Given the fact
that the DLA galaxy was already confirmed, it provided an ideal target for
PMAS in order to investigate the limitations of the instrument for observing
faint targets. We obtained a total of 2 hours integrations under good seeing
conditions and low airmass in September 2002.  The data analysis has revealed
that a nebula of \lya\ emission at $z=3.1538$ surrounds the confirmed DLA
galaxy, see Fig.~\ref{fig:map} (Christensen et al. 2004).
The spectra of the QSO and the nebula are shown in Fig.~\ref{fig:spec},
indicating the correlation between the position of the DLA line in the
spectrum and the \lya-emission line from the extended nebula.  We find that
the nebula has an extension of 3\arcsec$\times$5\arcsec, corresponding to
$\sim20\times40 $ kpc at the redshift of the absorber, and the total flux from
the source is $\sim$4 times larger than found in the discovery long slit
spectrum. Clearly, for extended objects the integral field spectra have
advantages over long slit spectroscopy. In this particular case we suspect
that the extended nebula has some substructure, so the flux derived from
standard slit spectroscopy is very dependent on the placement of the slit.

We find the total line flux $f_{\mathrm{\lya}}=2.8\times10^{-16}$
erg~cm$^{-2}$~s$^{-1}$ from the extended \lya\ emission line region from the
DLA galaxy. As argued in Christensen et al. (2004), a possible cause for this
extended emission is a galactic outflow driven by star formation and supernova
explosions in the central galaxy.  Analyses of the absorption line profiles in
Lu, Sargent and Barlow (1997) suggested that the DLA galaxy was a massive
rotating disk.  Based on the offset between the emission and the absorption
redshift (209~km~s$^{-1}$) of the \lya\ line in the long-slit spectrum the
presence of a disk was suggested by Djorgovski et al. (1996). We find a
slightly larger offset in our spectra.  Although the individual spectra are
noisy, we do not see any evidence in the integral field spectra for a large
scale rotation expected in the case of a galactic disk.

\section{Summary and future outlook}
Our pilot study has shown that using the PMAS integral field instrument we are
able to detect the absorbing galaxies and determine their redshifts for a few
known systems.

At low redshifts the impact parameter can very well be larger than the field of
view of PMAS, while at $z>2$ the systems confirmed to date have smaller impact
parameters $b<3\arcsec$ (Fynbo, M{\o}ller \& Warren 1999).

We are carrying on the project of identifying DLA galaxies specifically at
high redshift where many ($\sim$150) still await confirmation.  Possibly the
confirmed DLA galaxies at these redshifts are simply in the bright end of the
distribution. Low star formation rates for the DLA galaxies have been inferred
from the small metallicity enrichment with redshift (Pettini et al., 1999). A
low SFR results in a small \lya\ flux, e.g. a star formation rate of
1~M$_{\odot}$ yr$^{-1}$ gives rise to a \lya\ line flux of $< 10^{-17}$ \ecs\ 
at high redshifts assuming no dust obscuration. Additionally, the escape
fraction of the resonant \lya\ photons is highly dependent on the velocity
structure of the environment. Another explanation is that emission from DLA
galaxies are not detected due to a less fortunate placement of the slits.

For the Q2233+131 DLA absorber we have found evidence for extended \lya\ 
emission surrounding the DLA galaxy. Additionally, we found that the \lya\ 
emission is unlikely associated with a rotating galactic disk, which was
indicated by previous data. For this particular DLA galaxy we have found that
the detected flux using a 1\arcsec\ slit is a factor of 4 smaller than with
integral field spectroscopy, showing that integral field spectroscopy is an
efficient tool for this investigation.

\acknowledgements L.~Christensen acknowledges support by the German
Verbundforschung associated with the ULTROS project, grant no. 05AE2BAA/4, and
S.F.~S\'anchez acknowledges the support from the Euro3D Research Training
Network.


\begin{thebibliography}{}
\bibitem{beck} Becker, T.: 2002, Ph.D. thesis, Astrophysikalisches Institut Potsdam, Germany
\bibitem{} Bergeron, J., Boulade, O., Kunth, D., Tytler, D., Boksenberg, A., \& Vigroux, L.: 1988, A\&A, 191, 1 
\bibitem{} Chen, H. \& Lanzetta, K.~M.: 2003, ApJ, 597, 706
\bibitem{} Christensen, L.,  S\'anchez, S.~F., Jahnke, K., Becker, T., Kelz,
  A., Wisotzki, L., Roth, M.~M.: 2004, A\&A, 417, 487
\bibitem{} Cohen, J.: 2001, AJ, 121, 1275
\bibitem{} Djorgovski, S.~G., Pahre, M.~A., Bechtold, J., \& Elston, R.: 1996,
  Nature, 382, 234
\bibitem{} Fynbo, J.~U., M{\o}ller, P., \& Warren, S.~J.\ 1999, MNRAS, 305,
  849
\bibitem{} Haehnelt, M.G.,  Steinmetz, M. \& Rauch, M.: 1998, ApJ, 495, 647
\bibitem{} Hunstead, R.~W., Fletcher, A.~B., \& Pettini, M.: 1990, ApJ, 356,
  23
\bibitem{} Le Brun, V., Bergeron, J., Boisse, P., Deharveng, J.M.: 1997, A\&A,
  321, 733
\bibitem{} Lu, L., Sargent, W.~L.~W., and Barlow, T.~A.: 1997, ApJ, 484, 131
\bibitem{} M{\o}ller, P., Warren, S.~J., Fall, S.~M., Fynbo, J.~U., \&
  Jakobsen, P.: 2002, ApJ, 574, 51
\bibitem{} Pettini, M., Ellison, S.~L., Steidel, C.~C., \& Bowen, D.~V.\ 1999,
 ApJ, 510, 576  
\bibitem{} Prochaska, J., {Gawiser}, E., {Wolfe}, A., {Castro}, S.,
  {Djorgovski}, S.~G.: 2003, ApJL, 595, L9
\bibitem{} Rao, S.M. \& Turnshek, D.A., 2000, ApJS, 130, 1
\bibitem{} Rao, S.M.,  Nestor, D.~B.,  Turnshek, D.A.,  Lane, W.M.,  Monier,
  E.M.,  Bergeron, J.: 2003, ApJ, 595, 94 
\bibitem{} Roth, M.M., Bauer, S., Diones, F., Fechner, T., Hahn,~T., Kelz,~A.,
  Paschke, J., Popow, E., Schmoll, J., Wolter, D., Laux, U., and Altmann, W.:
  2000, Proc. SPIE, 4008, 277
\bibitem{} Storrie-Lombardi, L. \& Wolfe, A.: 2000, ApJ, 543, 552
\bibitem{} Turnshek, D.~A., Rao, S., Nestor, D., Lane, W., Monier, E., Bergeron, J., \& Smette, A.: 2001, ApJ, 553, 288 
\bibitem{} Warren, S.J., M{\o}ller, P., Fall, P., Jakobsen, P.: 2001, MNRAS,
  326, 759
\bibitem{}Wolfe, A.~M., Turnshek, D.~A., Smith, H.~E., \& Cohen, R.~D.: 1986,
  ApJS, 61, 249
\end{thebibliography}
\end{document}